\documentclass{article}
\usepackage{amsmath}
\usepackage{graphicx}
\usepackage{amsfonts}
\usepackage{amssymb}
\usepackage{geometry}
\usepackage{float}
\usepackage{bm}
\usepackage{color}
\usepackage{cite}

\begin{document}

\title{Quantum corrections to the  thermodynamics of rotating charged BTZ black hole in gravity's rainbow.}

\author{
B. C. L\"{u}tf\"{u}o\u{g}lu\thanks{%
bclutfuoglu@akdeniz.edu.tr} \\
%EndAName
Department of Physics, Akdeniz University, Campus 07058, Antalya, Turkey, \\
Department of Physics, University of Hradec Kr\'{a}lov\'{e}, \\
Rokitansk\'{e}ho 62, 500 03 Hradec Kr\'{a}lov\'{e}, Czechia. 
\and B. Hamil%
\thanks{%
hamilbilel@gmail.com} \\
%EndAName
D\'{e}partement de TC de SNV, Universit\'{e} Hassiba Benbouali, Chlef,
Algeria. \and 
L. Dahbi\thanks{%
l.dahbi@ens-setif.dz} \\
%EndAName
Teacher Education College of Setif, Messaoud Zeghar, Algeria.}
\date{}
\maketitle

\begin{abstract}
In this manuscript we investigate the thermal properties of the rotating charged BTZ black hole under the {\color{red} gravity's rainbow (GR) and generalized uncertainty principle (GUP) formalism. At first, we study the GR-corrected thermal quantities according to the usual Heisenberg algebra.} Then, we consider a deformed algebra which leads to a change in the Heisenberg uncertainty principle, and compare the Hawking temperature, entropy, thermodynamical volume, pressure and heat capacity functions with the previous results. Thus, we understand and interpret the quantum effects on the BTZ black hole. 
\end{abstract}

\section{Introduction}

Black hole physics and thermodynamics is one of the most interesting and challenging topics in general relativity and modern cosmology. Historically, in the early 1970s, Bekenstein suggested that black hole entropy could be described as proportional to black hole area \cite{B,J}. In the classical approach accepted at the time, it was thought that black holes do not emit radiation, so Bekenstein's suggestion was met with skepticism. Just a year later, Hawking confirmed that a black hole could be taken as a black body, thus might emit radiation if the quantum effects are taken into account \cite{H}. This process, later called Hawking radiation, showed that a well-defined temperature could be defined for a black hole. Accordingly, the thermodynamic properties of a black hole, and especially its thermodynamic stability, have recently been discussed in interesting studies by many authors {\color{red}\cite{0011,20,ts01,ts02,ncl1, Mann}.}

Another great challenge of theoretical physics is deriving a well-defined unified theory from the theories of gravity and quantum. In order to achieve this goal, many quantum gravity theories, such as the string theory \cite{Amati}, noncommutative geometry \cite{Girelli}, loop quantum gravity \cite{Rovelli, Carlip} have been proposed. All these theories present some differences with respect to each other, so they have some advantages and disadvantages. But they also have some common properties, such as the existence of a minimum measurable length. However, scientists were confused by the existence of a minimum length on the Planck scale, since the Planck length scale is not a Lorentz invariant quantity. To resolve this contradiction, modification of the usual distribution relation in the special theory of relativity was proposed. The modified dispersion relation yielded a new theory called doubly special relativity (DSR),  which has two observer-independent scales namely, the velocity of light $c$ and the Planck energy $E_{P}$ (or the Planck length $\ell_{P}$) \cite{Amelino, Magueijo, Smolin, Alfaro, Sahlmann, Smolinn}. It is worth noting that at the limit where the minimum length disappears, the DSR theory reduces to standard special relativity. In 2004, Magueijo and Smolin formulated the DSR in the curved spacetimes and called the formalism as gravity's rainbow (GR) (In the literature, sometimes it is also called rainbow gravity (RG).) \cite{Magueijo}. The basic assumption of the GR stands on the fact that the energy of the test particle must also affect the geometry of spacetime. Therefore,  the spacetime background has to be represented with the parameter-dependent family of metrics that depend on the energy of the considered particle. This parameter dependency creates the rainbow of the metric and recently a lot of papers are published on the black hole thermodynamics by considering various GR {\color{red}\cite{Ling,Kim,Feng,Li,Rudra, Heydarzade,Eslam1, Eslam2, Eslam3}.}

The existence of the minimum measurable length in quantum mechanics can also be obtained by the deformation of the usual Heisenberg algebra \cite{Kempf,Scardigli}. Such a deformation also affects the Heisenberg uncertainty principle (HUP), and as a result, the generalized uncertainty principle (GUP) has to be used instead of the HUP.  Recently, with increasing interest, we observe that many papers examine the effect of the GUP formalism on black hole physics and thermodynamics {\color{red} \cite{0012,Nouicer2007, Myung2007, Gangopadhyay2014,18,19,cc12,cc13,cc14,cc15,cc16,cc17,cc18,cc19,cc20,cc21,cc22,cc23,cc24,cc25, Marcos1, Marcos2, Marcos3}.} One of the common features of these studies is the modification of the Hawking radiation. In {\color{red} \cite{Cavaglia,Ali, ncl2}} the authors discussed whether the black holes may or may not be detected in the LHC experiments and claimed the existence of a remnant can have important phenomenological consequences for the observation of black holes at the LHC experiments. In addition, in the above studies,  they showed that the entropy of the black hole drastically changes by getting a logarithmic correction term with a negative sign which is consistent with the string theory and loop quantum gravity.

In 1992, Ba\~nados, Teiteloim and Zanelli (BTZ) considered a negative cosmological constant and studied the usual Einstein-Maxwell field equation in $2+1$ spacetime. They obtained a vacuum solution and interpreted them as the BTZ black hole solution which shows similarities to the $3+1$ dimensional Schwarzschild and Kerr black hole solutions \cite{BTZ}.  One year later, Ach\'ucarro and Ortiz discussed the rotating charged BTZ black hole solution \cite{rcBTZ}. Certain aspects of the thermodynamic geometry of rotating BTZ black holes are investigated for uncharged and charged cases, respectively in {\color{red}\cite{sarkar, akbar1, Eslam4, Eslam5}.} Recently, Alsaleh studied the thermodynamics of neutral BTZ black holes in the GR \cite{Salwa}. As it is mentioned above, the GUP formalism affects the thermodynamics of a black hole drastically. {\color{red}To this end, Iorio et al. considered the BTZ black hole in the GUP formalism and presented its impact on the usual thermodynamic quantities \cite{Iorio}.} To our best knowledge, an investigation on the thermodynamic features of the charged BTZ black holes in the GR formalism under the GUP has not been carried on before. This fact is the one of the main motivation of this manuscript. We construct the manuscript as follows: In Sec. \ref{sec2}, we consider a charged BTZ black hole and study its thermal quantities in GR formalism.  After analyzing the Hawking temperature, entropy, thermodynamical volume, Helmholtz free energy, pressure, internal energy and heat capacity functions, in Sec. \ref{sec3} we introduce the GUP formalism and we repeat a similar analyze in the GUP scenario. Finally, we give a brief conclusion in the last section.

\section{Thermodynamics of BTZ black holes under the effects of the GR} \label{sec2}

We consider the Hilbert action in $(2+1)$ Anti-de Sitter (AdS) spacetime with the electromagnetic field coupling \cite{1}%
\begin{equation}
\mathcal{I}=\int d^{3}x\sqrt{-g}\left( R-2\Lambda -\frac{1}{4}F_{ab}F^{ab}\right); \text{ \  \ }(a,b=0;1;2).
\end{equation}
Here, $R$ and $\Lambda$ represent the curvature scalar and cosmological constant where $\Lambda =-\ell ^{-2}<0$, while $\ell$ is the AdS radius.  The Einstein field equations that are expressed with 
the Einstein tensor, $G_{ab}$,  and stress-energy tensor, $T_{ab}$, in the form of
\begin{equation}
G_{ab}+\Lambda g_{ab}=\pi T_{ab},  \label{1}
\end{equation}%
gives the BTZ black hole solution via the given metric \cite{2}  
\begin{equation}
ds^{2}=-\mathcal{F}\left( r\right) dt^{2}+\frac{1}{\mathcal{F}\left(
r\right) }dr^{2}+r^{2}\left( d\phi -\frac{J}{2r^{2}}dt\right) ^{2},
\end{equation}%
where the lapse function, $\mathcal{F}\left( r\right)$, is taken as
\begin{equation}
\mathcal{F}\left( r\right) =-M+\frac{r^{2}}{\ell ^{2}}+\frac{J^{2}}{4r^{2}}%
-2Q^{2}\ln \frac{r}{\ell }.
\end{equation}
Here, three integration constants: $M$, $J$, and $Q$ denote mass, angular momentum (spin), and the charge of the BTZ black
hole, respectively. {\color{red} The roots of the lapse function are the horizons of the black hole. However, it is not easy to determine a precise expression of the horizons of the BTZ black hole. With the help of extremal points of the lapse function, $\frac{d\mathcal{F}(r)}{dr}\big|_{r=r_{ext}}=0$, one can obtain a condition, $\mathcal{F}(r_{ext})<0$, that ensures real-valued event horizons \cite{Akbar}.
\begin{eqnarray}
M> \frac{Q^2 }{2} \left(1 + \sqrt{1+ \frac{J^2}{Q^4l^2 }}\right)+ \frac{J^2}{2 \ell^2 Q^2 \left(1 + \sqrt{1+ \frac{J^2}{Q^4l^2 }}\right)}-2Q^2 \ln \sqrt{\frac{Q^2 }{2} \left(1 + \sqrt{1+ \frac{J^2}{Q^4l^2 }}\right)}. \label{newcond}
\end{eqnarray}
In a particular case, namely for $Q=0$ and $J\neq 0$, it reduces to 
$M> \frac{|J|}{\ell}$, while in another case, for $Q\neq 0$ and $J= 0$, it becomes $M>Q^2(1-2\ln Q)$. }
Considering the vanishing value of the lapse function, we express the mass of the black hole in terms of the outer event horizon, $r_{+}$, as
\begin{equation}
M=\frac{r_{+}^{2}}{\ell ^{2}}+\frac{J^{2}}{4r_{+}^{2}}-2Q^{2}\ln \frac{r_{+}}{\ell }. \label{m}
\end{equation}
To examine the GR effects, we follow the method given in \cite{3,4,5,6}, and modify the metric via $dt\rightarrow \frac{dt}{f\left( \frac{E}{E_{P}}\right) }$ and  $dx^{i}\rightarrow \frac{dx^{i}}{g\left( \frac{E}{E_{P}}\right) }$ by employing two arbitrary rainbow functions. So that, we obtain the GR modified metric in the following form
\begin{equation}
ds^{2}=-\mathcal{A}\left( r\right) dt^{2}+\frac{1}{\mathcal{B}\left(
r\right) }dr^{2}+\frac{r^{2}}{g^{2}\left( \frac{E}{E_{P}}\right) }\left(
d\phi -\frac{J}{2r^{2}}dt\right) ^{2},
\end{equation}%
where%
\begin{equation}
\mathcal{A}\left( r\right) =\frac{\mathcal{F}\left( r\right) }{f^{2}\left( 
\frac{E}{E_{P}}\right) };\text{ \  \  \  \ }\mathcal{B}\left( r\right)
=g^{2}\left( \frac{E}{E_{P}}\right) \mathcal{F}\left( r\right) .  \label{2}
\end{equation}%
Here, $E$ illustrates the energy at which the geometry is probed, while $E_{P}$ denotes the Planck energy. Then, we constraint the arbitrariness of the rainbow functions by choosing one of
the most interesting forms of the rainbow functions 
\begin{equation}
f\left( \frac{E}{E_{P}}\right) =1;\text{ \ }g\left( \frac{E}{E_{P}}\right) =%
\sqrt{1-\eta \left( \frac{E}{E_{P}}\right)^2}.  \label{3}
\end{equation}%
It is worth noting that this selection of rainbow functions has a physical background, such as its use in the string theory \cite{3,7}, quantum cosmology \cite{8},  loop quantum gravity \cite{9}, and $\kappa-$ Minkowski noncommutative spacetime \cite{10}. The assumption of $f\left( \frac{E}{E_{P}}\right) =1$, yields a time-like Killing vector in the GR as usual, so that, the local thermodynamic energy does not show a dependency on the energy of the test particle. 

First, we intend to derive the Hawking temperature of BTZ black hole in the presence of the GR. Following \cite{3}, we employ the formula 
\begin{equation}
T_{H}=\frac{1}{2\pi }\left. \sqrt{\frac{\partial \mathcal{A}\left( r\right) 
}{\partial r}\frac{\partial \mathcal{B}\left( r\right) }{\partial r}}\right
\vert _{r=r_{+}},  \label{4}
\end{equation}
and straightforwardly, we find the Hawking temperature in the GR framework in the form of 
\begin{equation}
T_{H_{GR-HUP}}=\frac{r_{+}}{2\pi }\left( \frac{2}{\ell ^{2}}-\frac{J^{2}}{2r_{+}^{4}}%
-\frac{2Q^{2}}{r_{+}^2}\right) \sqrt{1-\frac{\eta }{E_{p}^{2} r_{+}^{2}}}.  \label{t}
\end{equation}
Since the Hawking temperature can only take real and positive values, we obtain two constraining conditions. One of them stems from the GR formalism and provides a relation between the horizon and the rainbow parameter, $\eta $, in the Planck energy scale.  
\begin{equation}
r_{+}\geq \frac{\sqrt{\eta }}{E_{p}}{\color{red}.} \label{kosul1}
\end{equation}%
The second condition is based on the fact that the temperature must take a positive value and establishes a relationship between charge, angular momentum, space-time radius and horizon. 
Therefore, we can interpret it  as an intrinsic physical condition.  
\begin{eqnarray}
r_+ &\geq& \ell \sqrt{\frac{Q^2 }{2} \left(1 + \sqrt{1+ \frac{J^2}{Q^4l^2 }}\right)}. \label{rplus}
\end{eqnarray}
{\color{red} In fact, under the hypothesis $Q=\eta=0$ the reality condition, $M> \frac{|J|}{\ell}$, guarantees positive temperature. When $Q\neq 0$, Eq. \eqref{rplus} becomes useful.}
We find it interesting to perform a deeper analysis on the unification of these two conditions since they are based on different facts. Assuming the first condition sets a lower limit to the horizon than the second condition, we substitute Eq. \eqref{kosul1} in Eq. \eqref{t} and we obtain a correlation between the GR's formalism and the BTZ black hole properties. 
\begin{eqnarray}
0<\eta< \frac{Q^2 \ell^2 E_p^2}{2}\left(1+ \sqrt{1+\frac{J^2}{\ell^2 Q^4}}\right). \label{lowerlimit}
\end{eqnarray}
for $\eta > 0$. On the other hand, the black hole horizon cannot be greater than the length of spacetime, {\color{red}$\ell>r_+$}. If we use this as an upper bound in Eq. \eqref{rplus}, we get the following relationship between the black hole parameters.
\begin{eqnarray}
\frac{1}{\ell}\left(2-2Q^2 -\frac{J^{2}}{2\ell^{2}}\right) &\geq& 0.
\end{eqnarray}
Since $\ell$ has a non-zero positive value, we arrive at a new condition among the parameters as follows
\begin{eqnarray}
\ell &\geq& \frac{J}{2\sqrt{1-Q^2}}. \label{upperlimit}
\end{eqnarray}
For the following two sub-cases , where the BTZ black hole is considered as {\color{red} non-static uncharged} or {\color{red} static charged}, the conditions given in Eqs. \eqref{lowerlimit} and \eqref{upperlimit} reduce to  {\color{red}$0<\eta< \frac{|J| \ell E_p^2}{2}$}, \,\, $\ell >\frac{J}{2}$, and   $0<\eta< \frac{Q^2 \ell^2 E_p^2}{2}$,  \,\,    $\ell > 0$, respectively.

\begin{figure}[htbp]
\centering
\includegraphics[scale=1]{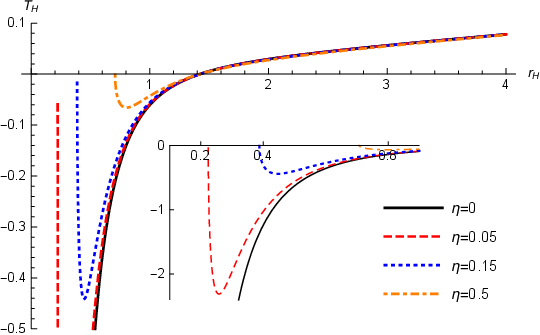}
\caption{Hawking temperature versus event horizon in the GR formalism for $J=1$, $Q=0.05$, $l=4$, $E_p=1$.} \label{fig1}
\end{figure}

According to this analysis, we plot the Hawking temperature versus the horizon in Fig. \ref{fig1} with the parameters that obey Eqs. \eqref{lowerlimit} and \eqref{upperlimit}. We observe that the GR formalism leads to a lower bound on the horizon, $r_{min}$, as predicted. Although it modifies the Hawking temperature, we see that these effects are dominant only in the nonphysical region, $r_{min} \leq r_+ \leq r_{phys}$, where the horizon is smaller than $1.42$ for the chosen representative parameter values. In the range $ r_{phys}\leq r_+ \leq r_{max}$, the GR corrected Hawking temperature mimics the usual one. 

\begin{figure}[htbp]
\centering
\includegraphics[scale=1]{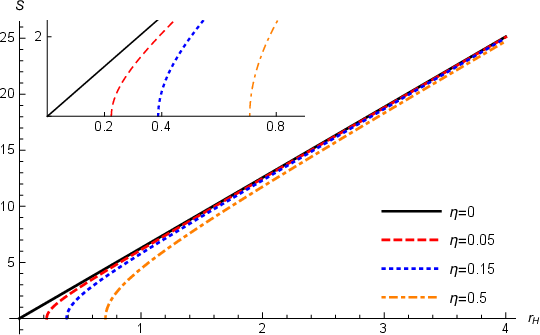}
\caption{Bekenstein entropy versus event horizon in the GR formalism for $J=1$, $Q=0.05$, $l=4$, $E_p=1$} \label{fig2}
\end{figure}

Next, we employ another fundamental postulate of the black hole thermodynamics to derive the
Bekenstein entropy, $S=\int \frac{dM}{T_{H}}$. By using Eqs. \eqref{m} and \eqref{t}, we find
\begin{equation}
S_{GR-HUP}=2\pi \int \frac{dr_{+}}{\sqrt{1-\frac{\eta }{%
E_{p}^{2}r_{+}^{2}}}}=2\pi r_{+}\sqrt{1-\frac{\eta }{E_{p}^{2}r_{+}^{2}}}.  \label{s}
\end{equation}%
We see that the Bekenstein entropy depends on the rainbow parameter. {\color{red} In other words, Eq. \eqref{s} shows that entropy does not
obey area law if $\eta\neq 0$. But when $\eta=0$, the area law for the BTZ black hole is clearly obeyed} \cite{11,12,13,14,15,16,17}. We depict GR corrected entropy function versus the horizon in Fig. \ref{fig2}. 

\noindent We observe a linear increase in entropy function for $\eta=0$. In the presence of the GR formalism, this linearity breaks down drastically especially at small horizon values i.e. for $r_H< (r_{phys}=1.42)$, where the BTZ black hole is unstable, hence, its entropy is not physically meaningful. In the range $r_{phys}\leq r_+ \leq r_{max}$, the GR corrected entropy differs from the usual one by the factor $\frac{\eta}{E_p^2 \ell^2}$.

\begin{figure}[htbp]
\centering
\includegraphics[scale=1]{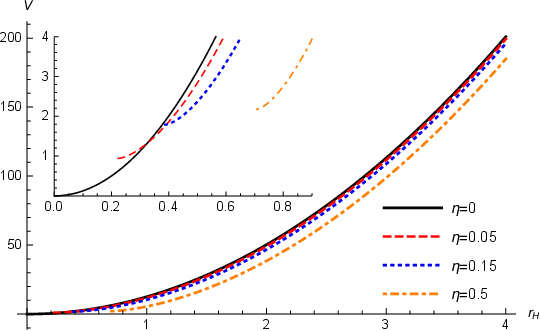}
\caption{Thermodynamical volume of the rotating charged BTZ black hole versus event horizon in the GR formalism for   $J=1$, $Q=0.05$, $l=4$, $E_p=1$.} \label{fig3}
\end{figure}

\noindent Since the entropy of the considered black hole is not proportional to the event horizon area, we would like to calculate its volume via the following formula
\begin{equation}
V=4\int Sdr_{+}.
\end{equation}%
We find %
{\color{red}
\begin{eqnarray}
V_{GR-HUP}&=&4\pi r_{+}^{2}\left \{ \sqrt{1-\frac{\eta }{E_{p}^{2}r_{+}^{2}}}-\frac{%
\eta }{E_{p}^{2}r_{+}^{2}}\ln \left[\frac{r_{+}}{\ell} \left(1+\sqrt{1-\frac{\eta }{%
E_{p}^{2}r_{+}^{2}}}
\right) \right] \right \} . \label{vol1}
\end{eqnarray}}
In the absence of GR formalism, we obtain $V=4\pi r_+^2$, which is the usual form of the  rotating charged BTZ black hole volume. In that case for $r_+=0$, volume becomes zero. However, in the GR formalism, a non-zero minimal volume value arises at minimal event horizon value in the form of
{\color{red}
\begin{eqnarray}
V_{min}= - \frac{4\pi\eta}{E_p^2}\ln \Bigg[ \sqrt{\frac{\eta}{E_p^2 \ell^2}}
\Bigg].
\end{eqnarray}}
We present the variation of the GR corrected black hole volume versus the event horizon in Fig. \ref{fig3}. We see that for $\eta=0.05$, $\eta=0.15$ and $\eta=0.50$, minimum horizon reads  $0.22$, $0.39$, $0.71$,  and accordingly minimum volume values are $0.94$, $1.79$,
$2.18$. In the stable region, we observe that the GR corrected volume value is smaller than the usual one. This deviation becomes greater for large GR parameters. 

\begin{figure}[htbp]
\centering
\includegraphics[scale=1]{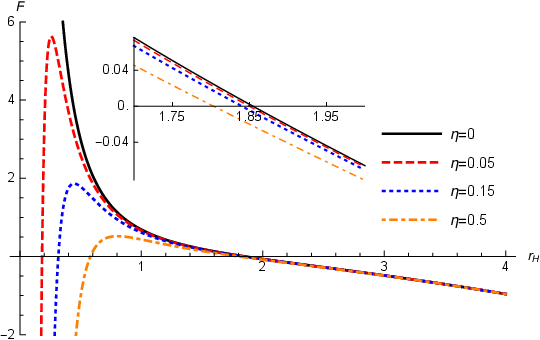}
\caption{Helmholtz free energy versus event horizon in the GR formalism for $J=1$, $Q=0.05$, $l=4$, $E_p=1$.} \label{fig4}
\end{figure}

\noindent Then, we investigate the Helmholtz free energy. For the derivation we use the well-known formula $F=-\int SdT$. By substituting  Eqs. \eqref{s} and \eqref{t}, we obtain the GR corrected Helmholtz free energy of the charged rotating BTZ black hole in terms of horizon as follows: 
{\color{red}
\begin{eqnarray}
F_{GR-HUP}=- \left( \frac{r_{+}^{2}}{\ell ^{2}}-\frac{3J^{2}}{4r_{+}^{2}}%
+2Q^{2}\ln \frac{r_{+}}{\ell}\right) -\frac{\eta }{E_{p}^{2}r_{+}^{2}}\left( \frac{J^{2}}{%
2r_{+}^{2}}+2Q^{2}\right) . \label{HFE1}
\end{eqnarray}}
We notice that the second term  represents the GR contribution. To illustrate the modification of the Helmholtz free energy, we depict it in terms of horizon in the presence and absence of the GR in Fig. \ref{fig4}.

We observe that in the absence of GR parameter, Helmholtz free energy decreases monotonically. In the presence of the GR formalism, it exhibits a different characteristic behavior in the horizon range where the black hole is unstable. On the contrary, we observe that when the black hole becomes stable, the GR effect does not make a significant decrease on the thermodynamic function.    

Next, we examine the thermodynamic pressure of the black hole, defined as proportional to the first derivative of the Helmholtz free energy with respect to the volume. 
\begin{equation}
P=-\frac{dF}{dV}. 
\end{equation}%
We obtain the pressure in the form of
\begin{eqnarray}
P_{GR-HUP}=\frac{1}{8\pi \sqrt{1-\frac{\eta }{%
E_{p}^{2}r_{+}^{2}}}}\left[ \left( \frac{2}{\ell ^{2}}+\frac{%
3J^{2}}{2r_{+}^{4}}+\frac{2Q^{2}}{r_{+}^2}\right) -\frac{4 \eta }{E_{p}^{2}r_{+}^{2}}\left( \frac{J^{2}}{%
2r_{+}^{4}}+\frac{Q^{2}}{r_{+}^2}\right) %
\right]. \label{pres1}
\end{eqnarray}%

\begin{figure}[htbp]
\centering
\includegraphics[scale=1]{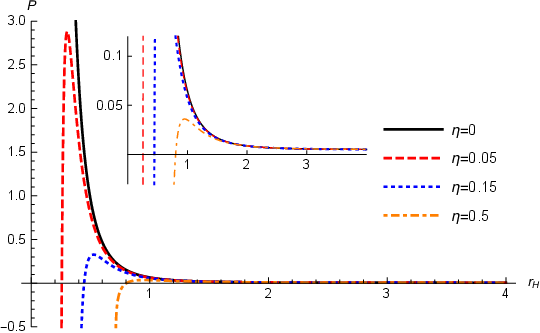}
\caption{Pressure versus event horizon in the GR formalism for $J=1$, $Q=0.05$, $l=4$, $E_p=1$.} \label{fig5}
\end{figure}

\noindent In the usual case, where $\eta=0$, we see that the pressure decreases only by taking positive values and converges to a certain value that depends on the characteristic properties of the black hole. Since the pressure has to be real valued, we see that RG correction enforces a lower limit bound on the event horizon. In order to give a detailed discussion, we plot the RG corrected pressure versus event horizon in Fig. \ref{fig5}. 

\noindent We see that negative pressure can arise only when the black hole is unstable. Moreover in that range the pressure first increase rapidly, and then decrease. We conclude that the GR correction is dominant only in this non physical range. On the other hand, in the stable range, GR correction does not change the pressure substantially. In greater horizon values, the pressure saturates at a non zero value.  

Next, we examine the internal energy using the following thermodynamic relations, 
\begin{equation}
U=\int TdS,
\end{equation}%
Considering Eqs. \eqref{t} and \eqref{s}, we obtain the internal energy independent from the effect of GR as follows:{\color{red}
\begin{equation}
U_{GR-HUP}=\frac{r_{+}^{2}}{\ell ^{2}}+\frac{J^{2}}{4r_{+}^{2}}-2Q^{2}\ln\frac{r_{+}}{\ell} ,
\end{equation}}

\begin{figure}[htbp]
\centering
\includegraphics[scale=1]{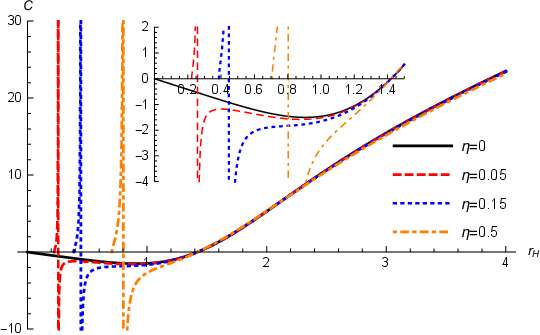}
\caption{Heat capacity versus event horizon in the GR formalism for $J=1$, $Q=0.05$, $l=4$, $E_p=1$.} \label{fig6}
\end{figure}

\noindent We see that the internal energy is not modified by the GR formalism. Then, we discuss the stability of the GR corrected BTZ black hole. To this end, we derive the heat capacity and analyze its behaviour. According to the following definition
\begin{equation}
C_{J,Q}=T_{H}\left( \frac{\partial S}{\partial T_{H}}\right) _{J,Q},
\label{c}
\end{equation}%
we express the heat capacity of the black hole with the help of Eqs. (\ref{s}), (\ref{t}), and
(\ref{c}) in the form of 
\begin{eqnarray}
C_{J,Q_{GR-HUP}}=2\pi r_{+} \left[\frac{\left(
\frac{2}{\ell^{2}}-\frac{J^{2}}{2r_{+}^{4}}-\frac{2Q^{2}}{r_{+}^2}\right)}{\left(
\frac{2}{\ell^{2}}+\frac{3J^{2}}{2r_{+}^{4}}+\frac{2Q^{2}}{r_{+}^2}\right)-\left(\frac{2J^{2}}{r_{+}^{4}}+\frac{4Q^{2}}{r_{+}^2}\right)\frac{\eta }{E_{p}^{2}r_{+}^{2}}}\right]\sqrt{1-\frac{\eta }{E_{p}^{2}r_{+}^{2}}}.
 \label{cc}
\end{eqnarray}%
This corrected heat capacity function reduces to standard condition for $\eta =0$. We depict GR corrected heat capacity versus horizon in Fig. \ref{fig6}. 

\begin{figure}[htbp]
\centering
\includegraphics[scale=0.75]{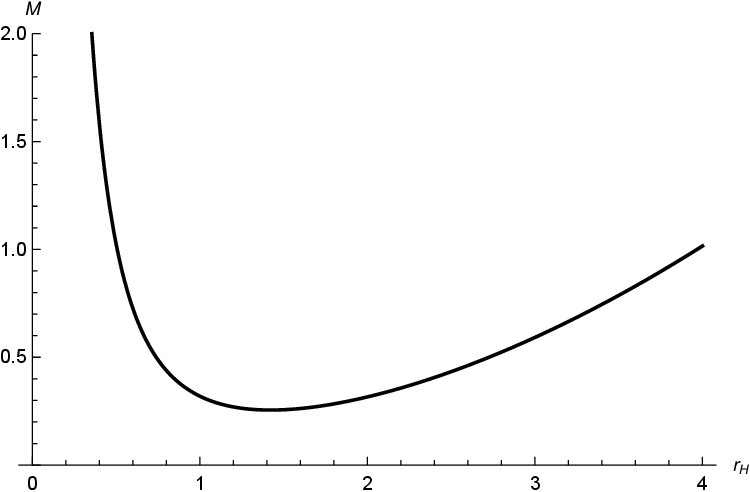}
\caption{{\color{red}Mass} function versus event horizon in the GR formalism for $J=1$, $Q=0.05$, $l=4$, $E_p=1$.} \label{fig8}
\end{figure}

\noindent We observe that in the nonphysical range black hole is unstable, When $r_H > r_{phys}$, black hole becomes stable. For $r_H = r_{phys}$, black hole with the mass
\begin{eqnarray}
M&=& Q^2\Bigg[\sqrt{1+ \frac{J^2}{\ell^2 Q^4}}+ \ln 2- \ln \bigg(Q^2\left(1+\sqrt{1+ \frac{J^2}{\ell^2 Q^4}}\right)\bigg) \Bigg], \label{remnant}
\end{eqnarray}
does not radiate. This mass value corresponds the minimal value of the mass function given in Fig. \ref{fig8}.

\section{Quantum corrections to the  thermodynamics of the BTZ black hole under the rainbow gravity}\label{sec3}

\begin{figure}[htbp]
\centering
\includegraphics[scale=1]{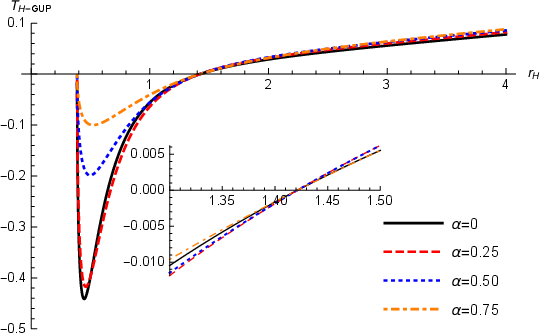}
\caption{GR-corrected Hawking temperature versus event horizon in the GUP formalism for $\eta=0.15$, $J=1$, $Q=0.05$, $l=4$, $E_p=1$.} \label{fig11}
\end{figure}

In this section, we consider a deformed Heisenberg algebra and examine the quantum effects on the thermodynamics of BTZ black hole in the presence of RG formalism. 
To this end, we start by introducing the considered deformed algebra which leads to the following GUP \cite{24,25}: 
\begin{equation}
\Delta P\Delta X\geq \left( 1-\alpha \left( \Delta P\right) +\alpha
^{2}\left( \Delta P\right) ^{2}\right) ,  \label{7}
\end{equation}%
where $\alpha $ is a positive parameter. {\color{red} It is worth noting that there are other scenarios where the deformation parameter is addressed with a negative quantity \cite{Jizba,  Buoninfante}. Besides,  there are also cases where the deformation parameter is considered as a dynamic variable rather than a constant, leading to a more general perspective and analysis \cite{Chen, Petruzziello}. We also would like to emphasize that GUP can be formulated in a completely different way, particularly in the presence of the BTZ black hole. For example, Iorio et al used the form that gives $\Delta P$  to the power of $3/2$ in \cite{Iorio}. Now, let us continue by rewriting} Eq. \eqref{7} as 
\begin{equation}
\Delta P\geq \frac{\left( \Delta X+\alpha \right) }{2\alpha ^{2}}\left( 1-%
\sqrt{1-\frac{4\alpha ^{2}}{\left( \Delta X+\alpha \right) ^{2}}}\right) ,
\end{equation}%
and Taylor expand. We get
\begin{equation}
\Delta P\geq \frac{1}{\left( \Delta X\right) }\left( 1-\frac{\alpha }{\left(
\Delta X\right) }+\frac{2\alpha ^{2}}{\left( \Delta X\right) ^{2}}+\mathcal{O%
}\left( \alpha ^{3}\right) \right) .  \label{8}
\end{equation}%
{\color{red}Next,} we use the saturated form of the uncertainty principle $E\Delta X\geq 1$, which follows from the saturated form of the HUP, $\Delta P\Delta X\geq 1$ \cite{26}, in Eq. \eqref{8}. We get
\begin{equation}
E_{GUP}\geq E\left( 1-\frac{\alpha }{\left( \Delta X\right) }+\frac{2\alpha
^{2}}{\left( \Delta X\right) ^{2}}+\mathcal{O}\left( \alpha ^{3}\right)
\right) .
\end{equation}%
Here, $E$ is the energy of the tunneling particles and $E_{GUP}$ is the corrected energy of them. For a particle with corrected energy, the tunneling probability of crossing the event horizon is \cite{25}:
\begin{equation}
\Gamma \simeq \exp \left( \frac{-2\pi E_{GUP}}{\kappa }\right) .  \label{9}
\end{equation}%
Then, we can compare Eq. \eqref{9} with the Boltzmann distribution,  $\exp \left( -%
\frac{E}{T}\right)$, and find the quantum-corrected Hawking temperature of particles with energy $E$:
\begin{equation}
T_{H_{GR-GUP}}=\frac{T_H}{\left( 1-\frac{\alpha }{r_{+}}+\frac{2\alpha ^{2}}{r_{+}^{2}}%
\right)}. \label{tgup}
\end{equation}
We note that when $\alpha=0$ Eq. \eqref{tgup} reduces to Eq. \eqref{t}.  For small $\alpha$ values the GUP-corrected Hawking temperature reads
\begin{equation}
T_{H_{GR-GUP}}\simeq T_H\left[ 1+\frac{\alpha }{r_{+}} 
-\frac{\alpha ^{2}}{r_{+}^{2}}+\mathcal{O}(\alpha^3)
\right].
\end{equation}
We depict GR-GUP corrected Hawking temperature versus horizon in Fig. \ref{fig11}.  We observe that quantum corrections increase the Hawking temperature in physical region.  

Next, study the entropy function. We obtain the GR-GUP corrected entropy up to the second order correction as 
{\color{red}
\begin{equation}
S_{GR-GUP}=2\pi r_{+}\sqrt{1-\frac{\eta }{E_{p}^{2}r_{+}^{2}}}-2\pi \alpha
\ln \left[ \frac{r_{+}}{\ell}\left(1+\sqrt{1-\frac{\eta }{%
E_{p}^{2}r_{+}^{2}}}
\right) \right] - \frac{4\pi \alpha ^{2}}{\sqrt{\eta}}\arctan \sqrt{\frac{{\eta}}{{{E_{p}^{2}r_{+}^{2}}-\eta}}}.  \label{10}
\end{equation}}

\begin{figure}[htbp]
\centering
\includegraphics[scale=1]{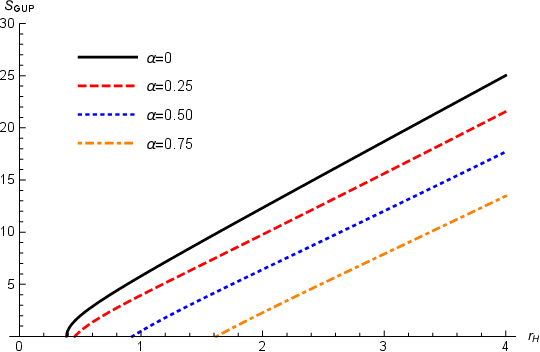}
\caption{GR-corrected entropy versus event horizon in the GUP formalism for \,\, $\eta=0.15$, $J=1$, $Q=0.05$, $l=4$, $E_p=1$.} \label{fig12}
\end{figure}

\noindent We see that the GUP correction terms leads to a decrease in the  entropy function.  In the absence of GUP parameter Eq. \eqref{10} reduces to Eq. \eqref{s}.   We demonstrate the behaviour of the entropy function versus event horizon in Fig. \ref{fig12}. 

\noindent We observe that entropy decreases more at greater GUP-parameter  value.  We also notice that the quantum corrected entropy of the BTZ black hole in gravity's rainbow in tunneling formalism, is of the form of
the predicted by string theory and loop quantum gravity.

Next, we investigate the thermodynamical volume of the black hole for a given entropy. We find
{\color{red}
\begin{eqnarray}
V_{GR-GUP} &=&4\pi r_{+}^{2}\left[\left( 1-\frac{2\alpha }{r_{+}}\right) \sqrt{1-\frac{\eta 
}{E_{p}^{2}r_{+}^{2}}}-\left( \frac{\eta }{E_{p}^{2}r_{+}^{2}}+\frac{\alpha }{r_{+}%
}+\frac{4\alpha ^{2}}{r_{+}^{2}}\right) \ln \left[\frac{r_{+}}{\ell}\left(1+\sqrt{1-\frac{\eta }{E_{p}^{2}r_{+}^{2}}}
\right) \right]\right] \nonumber \\
&-&  \frac{16 \pi r_+ \alpha ^{2}}{\sqrt{\eta}}
\arctan \sqrt{\frac{{\eta}}{{{E_{p}^{2}r_{+}^{2}}-\eta}}} . \label{VGUP}
\end{eqnarray}}

\begin{figure}[htbp]
\centering
\includegraphics[scale=1]{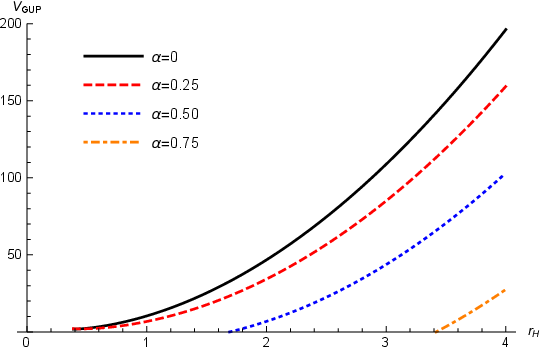}
\caption{GR-corrected volume versus event horizon in the GUP formalism for $\eta=0.15$, $J=1$, $Q=0.05$, $l=4$, $E_p=1$.} \label{fig13}
\end{figure}

\noindent In the HUP limit, Eq. \eqref{VGUP} becomes identical to Eq. \eqref{vol1}. We present the plot of thermodynamical volume versus event horizon in Fig. \ref{fig13}. We observe that with greater alpha parameters, smaller volumes are formed on the same horizon. 

\begin{figure}[htbp]
\centering
\includegraphics[scale=1]{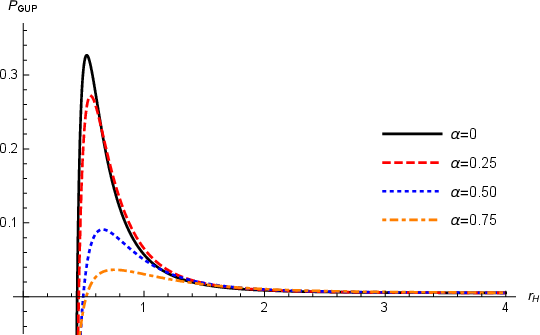}
\caption{GR-corrected pressure versus event horizon in the GUP formalism for  $\eta=0.15$, $J=1$, $Q=0.05$, $l=4$, $E_p=1$.} \label{fig15}
\end{figure}

Next, we investigate the quantum corrected pressure function. We find
\begin{eqnarray}
P_{GR-GUP}=\frac{1}{8\pi \sqrt{1-\frac{\eta
}{E_{p}^{2}}\frac{1}{r_{+}^{2}}}}\frac{\omega_1 \left( 1-\frac{\alpha }{r_{+}}+\frac{2\alpha ^{2}}{r_{+}^{2}}\right)
-\omega_2 \left(
\frac{\alpha }{r_{+}}-\frac{4\alpha ^{2}}{r_{+}^{2}}\right) }{ \left( 1-\frac{%
\alpha }{r_{+}}+\frac{2\alpha ^{2}}{r_{+}^{2}}\right) ^{2}}, \label{pres2}
\end{eqnarray}
where
\begin{subequations}
\begin{eqnarray}
\omega_1&=& \left( \frac{2}{\ell ^{2}}+\frac{3J^{2}}{%
2r_{+}^{4}}+\frac{2Q^{2}}{r_{+}^{2}}\right) -\left( \frac{2J^{2}}{r_{+}^{4}}+%
\frac{4Q^{2}}{r_{+}^{2}}\right) \frac{\eta }{E_{p}^{2}}\frac{1}{r_{+}^{2}} ,\\
\omega_2&=&\left( \frac{2}{\ell ^{2}}-\frac{J^{2}}{2r_{+}^{4}}-\frac{2Q^{2}}{r_{+}^{2}}%
\right) \left( 1-\frac{\eta }{E_{p}^{2}}\frac{1}{r_{+}^{2}}\right)  .
\end{eqnarray}
\end{subequations}
For $\alpha=0$, Eq. \eqref{pres2} reduces to Eq. \eqref{pres1}. We show the variation of the quantum corrected pressure with respect to the event horizon in Fig. \ref{fig15}. We notice that the quantum effects are relatively more effective in the non-physical region.

\begin{figure}[htbp]
\centering
\includegraphics[scale=1]{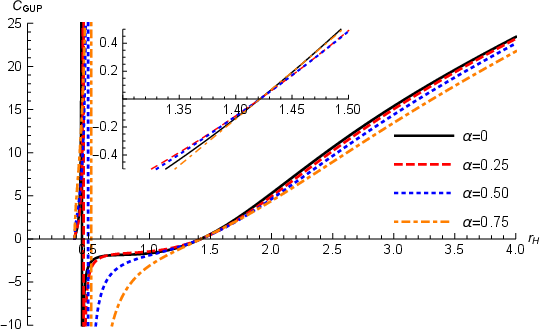}
\caption{GR-corrected entropy versus event horizon in the GUP formalism for $\eta=0.15$, $J=1$, $Q=0.05$, $l=4$, $E_p=1$.} \label{fig16}
\end{figure}

Finally, we investigate the quantum corrected heat capacity  function. We employ  Eq. (\ref{c}) for the derivation. In this scenario we find the heat capacity in the form of  
\begin{eqnarray}
C_{J,Q_{GR-GUP}}&=&2\pi r_{+} \left[\frac{\left( \frac{2}{\ell ^{2}}-\frac{J^{2}}{%
2r_{+}^{4}}-\frac{2Q^{2}}{r_{+}^2}\right) \left( 1-\frac{\alpha }{r_{+}}+\frac{2\alpha ^{2}}{r_{+}^{2}}\right)^2 }
{\omega_3-\omega_4\frac{\eta}{E_p^2r_{+}^2}}\right]\sqrt{1-\frac{\eta}{E_{p}^{2}r_{+}^{2}}}, \label{CGUP}
\end{eqnarray}
where 
\begin{subequations}
\begin{eqnarray}
\omega_3&=&\left( \frac{2}{\ell ^{2}}+\frac{3J^{2}}{2r_{+}^{4}}+\frac{2Q^{2}}{r_{+}^2}\right)\left( 1-\frac{\alpha }{r_{+}}+\frac{2\alpha ^{2}}{r_{+}^{2}}\right)-\left( \frac{2}{\ell ^{2}}-\frac{J^{2}}{2r_{+}^{4}}-\frac{2Q^{2}}{r_{+}^2}\right)\left(\frac{\alpha}{r_{+}}-\frac{4\alpha^2}{r_{+}^2}\right),\\
\omega_4&=&\left( \frac{2}{\ell ^{2}}+\frac{3J^{2}}{2r_{+}^{4}}+\frac{2Q^{2}}{r_{+}^2}\right)\left( 1-\frac{\alpha }{r_{+}}+\frac{2\alpha ^{2}}{r_{+}^{2}}\right)-\left( \frac{2}{\ell ^{2}}-\frac{J^{2}}{2r_{+}^{4}}-\frac{2Q^{2}}{r_{+}^2}\right)\left(1-\frac{2\alpha^2}{r_{+}^2}\right).
\end{eqnarray}
\end{subequations}
We see that in the absence of the quantum corrections, Eq. \eqref{CGUP} reduces to Eq. \eqref{cc}. We notice that the numerator is the product of two terms. However, the term that arises from the quantum correction does not have a real root. Therefore, the event horizon value, where the heat capacity is equal to zero, does not change. Thus, in the GUP case, when the black hole becomes stable its mass value is the same as it is in the HUP case. In Fig. \ref{fig16}, we give the variation of the heat capacity with respect to the event horizon in order to interpret what quantum corrections change. 

\noindent We observe that when the black hole becomes stable, the heat capacity function takes smaller values for greater quantum deformation parameter values due to quantum corrections. 

\section{Conclusions}
In the present manuscript, we consider a rotating charged BTZ black hole in gravity's rainbow {\color{red}(GR) under the generalized uncertainty principle (GUP) formalism. At first, in the Heisenberg uncertainty principle case, we show that the GR approach provides}  an extra constraint on the event horizon value. With a  detailed analyze, we investigate the Hawking temperature, entropy, thermodynamical volume, Helmholtz free energy, pressure, internal energy and heat capacity functions. We find that there is a physical and nonphysical region associated with the stability of the black hole.  Then, we consider a quantum deformation {\color{red} in the Heisenberg algebra, that leads to the GUP,} and examine the same functions according to the latter scenario. We find that quantum deformations increase the Hawking temperature, while they decrease the entropy, volume and  heat capacity in the physical region. We demonstrate all these effects on these functions with graphs.

\section*{Acknowledgments}
{\color{red} The authors thank the referees for a thorough reading of
our manuscript and for constructive suggestions.} One of the authors of this manuscript, BCL, is supported by the Internal  Project,  [2022/2218],  of  Excellent  Research  of  the  Faculty  of  Science  of Hradec Kr\'alov\'e University.


\begin{thebibliography}{99}
\bibitem{B} J. D. Bekenstein, Phys. Rev. D \textbf{7}, 2333 (1973).

\bibitem{J} J. D. Bekenstein, Phys. Rev. D \textbf{9}, 3292 (1974).

\bibitem{H} S. W. Hawking, Commun. Math. Phys. \textbf{43}, 199 (1975).

\bibitem{0011} R. J. Adler, P. Chen, D. I. Santiago, Gen. Rel. Grav. \textbf{33}, 2101 (2001).

\bibitem{20} B. Bolen, M. Cavagli\`{a}, Gen. Relativ. Gravit. \textbf{37},
1255 (2005).

{\color{red}
\bibitem{ncl1}F. Scardigli, Int. J. Geom. Meth. Mod. Phys. \textbf{17}, 2040004 (2020).

\bibitem{Mann} A. M. Frassino, R. B. Mann, J. R. Mureika, 
Phys. Rev. D \textbf{92}, 124069 (2015).}

\bibitem{ts01} V. G. Czinner, H. Iguchi, Phys. Lett. B \textbf{752}, 306 (2016).

\bibitem{ts02} M. S. Ma, R. Zhao, Y. Q. Ma, Gen. Relat. Grav. \textbf{49}, 79 (2017).

\bibitem{Amati} D. Amati, M. Ciafaloni, G. Veneziano, Phys. Lett. B \textbf{216}, 41
(1989).

\bibitem{Girelli} F. Girelli, E. R. Livine, D. Oriti, Nucl. Phys. B \textbf{%
708}, 411 (2005).

\bibitem{Rovelli} C. Rovelli, Living Rev. Rel. \textbf{1}, 1 (1998).

\bibitem{Carlip} S. Carlip, Rep. Prog. Phys. \textbf{64}, 885 (2001).

\bibitem{Amelino} G. Amelino-Camelia, Int. J. Mod. Phys. D \textbf{11}, 35
(2002).

\bibitem{Magueijo} J. Magueijo, L. Smolin, Class. Quantum Grav. \textbf{21},
1725 (2004).

\bibitem{Smolin} J. Magueijo, L. Smolin, Phys. Rev. Lett. \textbf{88},
190403 (2002).

\bibitem{Alfaro} J. Alfaro, H. A. Morales-Tecotl, L. F. Urrutia, Phys. Rev.
D \textbf{65}, 103509 (2002).

\bibitem{Sahlmann} H. Sahlmann, T. Thiemann, Class. Quantum Grav. \textbf{23}, 909 (2006).

\bibitem{Smolinn} L. Smolin, Nucl. Phys. B \textbf{742}, 142 (2006).

\bibitem{Ling} Y. Ling, X. Li, H. B. Zhang, Mod. Phys. Lett. A \textbf{22}%
, 2749 (2007)

\bibitem{Kim} Y. W. Kim, S. K. Kim, Y. J. Park, Eur. Phys. J. C \textbf{76%
}, 557 (2016).

\bibitem{Feng} Z. W. Feng, S. Z. Yang, Phys. Lett. B \textbf{772}, 737
(2017).

\bibitem{Li} H. Li, Y. Ling, X. Han, Class. Quantum Grav. \textbf{26},
065004 (2009).

\bibitem{Rudra} P. Rudra, M. Faizal, A. F. Ali, Nucl. Phys. B \textbf{909}, 725 (2016).

\bibitem{Heydarzade} Y. Heydarzade, P. Rudra, F. Darabi, A. F. Ali, M.
Faizal, Phys. Lett. B \textbf{774}, 46 (2017).

{\color{red} 
\bibitem{Eslam1} S. H. Hendi, S. Panahiyan, B. Eslam Panah, M. Momennia, Eur. Phys. J. C \textbf{76}, 150 (2016).

\bibitem{Eslam2} S. H. Hendi, B. Eslam Panah, S. Panahiyan,  Phys. Lett. B \textbf{769}, 191 (2017).

\bibitem{Eslam3}B. Eslam Panah, Phys. Lett. B \textbf{787}, 45 (2018).

}

\bibitem{Kempf} A. Kempf, G. Mangano, R. B. Mann, Phys. Rev. D \textbf{52}%
, 1108 (1995).

\bibitem{Scardigli} F. Scardigli, Phys. Lett. B \textbf{452}, 39 (1999).





\bibitem{Nouicer2007} K. Nouicer, Phys. Lett. B \textbf{646}, 63 (2007).

\bibitem{Myung2007} Y. S. Myung, Y. W. Kim, Y. J. Park, Phys. Lett. B \textbf{645}, 393 (2007).

\bibitem{Gangopadhyay2014} S. Gangopadhyay, A. Dutta, A. Saha, Gen. Relativ. Grav. \textbf{46}, 1661 (2014).


\bibitem{18} M. Faizal, M. M. Khalil, Int. J. Mod. Phys. A \textbf{30},
1550144 (2015).

\bibitem{0012} Y. C. Ong, JCAP \textbf{09}, 015 (2018).

\bibitem{cc12} H. Hassanabadi, E. Maghsoodi, W. S. Chung, Eur. Phys. J. C \textbf{79}, 358 (2019).

\bibitem{cc13} E. Maghsoodi, H. Hassanabadi, W. S. Chung, Prog. Theor. Exp. Phys. \textbf{2019}, 083E03 (2019).

\bibitem{cc14} H. Hassanabadi, E. Maghsoodi, W. S. Chung, M. de Montigny, Eur. Phys. J. C \textbf{79}, 936 (2019).

\bibitem{cc15} E. Maghsoodi, H. Hassanabadi, W. S. Chung, EPL \textbf{129}, 59001 (2020).

\bibitem{cc17} H. Hassanabadi, N. Farahani, W. S. Chung B. C. L\"utf\"uo\u{g}lu, EPL \textbf{130}, 40001 (2020).

\bibitem{cc18} N. Farahani, H. Hassanabadi, W. S. Chung, B. C. L\"utf\"uo\u{g}lu, S. Zarrinkamar, EPL \textbf{132}, 50009 (2020).

\bibitem{cc25} X. Guo, K. Liang, B. Mu, P. Wang, M. Yang, Eur. Phy. J. C \textbf{80}, 745 (2020)

\bibitem{cc16} H. Hassanabadi, W. S. Chung, B. C. L\"utf\"uo\u{g}lu, E. Maghsoodi, Int. J. Mod. Phys. A \textbf{36}, 2150036 (2021).


\bibitem{cc19} B. Hamil, B. C. L\"utf\"uo\u{g}lu,  EPL \textbf{133}, 30003 (2021).
\bibitem{cc20} B. Hamil, B. C. L\"utf\"uo\u{g}lu,  EPL \textbf{134}, 50007 (2021).
\bibitem{cc21} B. Hamil, B. C. L\"utf\"uo\u{g}lu,  EPL \textbf{135}, 59001 (2021).


\bibitem{19} L. Petruzziello, Class. Quantum Grav. \textbf{38}, 135005
(2021).

\bibitem{cc22} S. Hassanabadi, J. Kriz, W. S. Chung, B. C. L\"utf\"uo\u{g}lu, E. Maghsoodi,  H. Hassanabadi, Eur. Phys. J. Plus \textbf{136}, 918 (2021).

\bibitem{cc23} B. C. L\"utf\"uo\u{g}lu,  B. Hamil, L. Dahbi, Eur. Phys. J. Plus \textbf{136}, 976 (2021).

\bibitem{cc24} A. Bera, S. Dalui, S. Ghosh, E. C. Vagenas, arxiv 2109.00330v1 [gr-qc].


{\color{red}
\bibitem{Marcos1}
M. A. Anacleto, F. A. Brito, E. Passos, Phys. Lett. B \textbf{749} 181 (2015),


\bibitem{Marcos2}
M. A. Anacleto, F. A. Brito, B. R. Carvalho, A. G. Cavalcanti, E. Passos, J. Spinelly, Gen. Rel. Grav. \textbf{50}, 23 (2018).

\bibitem{Marcos3}
M. A. Anacleto, F. A. Brito, B. R. Carvalho, E. Passos, Adv. High Energy Phys. \textbf{2021}, 6633684 (2021).

}


\bibitem{Cavaglia} M. Cavaglia, S. Das, R. Maartens, Class. Quantum Grav. 
\textbf{20}, 205 (2003).

\bibitem{Ali} A. F. Ali, J. High Energy Phys. \textbf{1209}, 067 (2012).

{\color{red}
\bibitem{ncl2}F. Scardigli, Symmetry \textbf{12}, 1519 (2020).}

\bibitem{BTZ} M. Ba\~nados, C. Teitelboim, J. Zanelli, Phys. Rev. Lett. \textbf{69}, 1849 (1992).

\bibitem{rcBTZ} A. Ach\'ucarro, M. E. Ortiz, Phys. Rev. D \textbf{48}, 3600 (1993).

\bibitem{sarkar} T. Sarkar, G. Sengupta, B. N. Tiwari, J. High Energy Phys. \textbf{11}, 015 (2006).

\bibitem{akbar1} M. Akbar, H. Quevedo, K. Saifullah, A. Sanchez, S. Taj, Phys. Rev. D \textbf{83}, 084031 (2011).

{\color{red}
\bibitem{Eslam4} S. H. Hendi, S. Panahliyan, S. Upadhyay, B. Eslam Panah, Phys. Rev. D \textbf{95}, 084036 (2017).

\bibitem{Eslam5} B. Eslam Panah, S. Panahliyan, S. H. Hendi,  PTEP \textbf{2019}, 013E02 (2019).
}

\bibitem{Salwa} S. Alsaleh, Int. J. Mod. Phys. A \textbf{32}, 1750076 (2017).

{\color{red}
\bibitem{Iorio} A. Iorio, G. Lambiase, P. Pais, F. Scardigli, Phys. Rev. D \textbf{101}, 105002 (2020).
}


\bibitem{1} B. Pourhassan, A. \"{O}vg\"{u}n, \.{I}. Sakall\i , Int. J.
Geom. Meth. Mod. Phys. \textbf{17}, 2050156 (2020).

{\color{red} 
\bibitem{Akbar} M. Akbar, A. A.  Siddiqui, Phys. Lett. B \textbf{656}, 217 (2007).
} 

\bibitem{2} Y. -P. Zhang , S. -W. Wei, Y.-X. Liu, Phys. Lett. B 
\textbf{810}, 135788 (2020).

\bibitem{3} A. F. Ali, M. Faizal, M. M. Khalil, Nucl. Phys. B \textbf{%
894}, 341 (2015).

\bibitem{4} Z. -W. Feng, S. -Z. Yang, Phys. Lett. B \textbf{772}, 737 (2017).

\bibitem{5} S. H. Hendi, Gen. Relativ. Gravit. \textbf{48}, 1 (2016).

\bibitem{6} Z. Z. Ma, Phys. Lett. B \textbf{666}, 376 (2008).

\bibitem{7} G. Amelino-Camelia, J. R. Ellis, N. Mavromatos, D. V.
Nanopoulos, Int. J. Mod. Phys. A \textbf{12}, 607 (1997).

\bibitem{8} G. Amelino-Camelia, J. R. Ellis, N. Mavromatos, D. V.
Nanopoulos, S. Sarkar, Nature \textbf{393}, 763 (1998).

\bibitem{9} G. Amelino-Camelia, Living Rev. Rel. \textbf{16}, 5 (2013).

\bibitem{10} J. Lukierski, H. Ruegg, W. J. Zakrzewski, Ann. Phys.\textbf{%
\ 243}, 90 (1995).

\bibitem{11} A. Medved, E. C. Vagenas, Mod. Phys. Lett. A \textbf{20},
1723 (2005).

\bibitem{12} Y. -X. Chen, J. -L. Li, K. -N. Shao, Europhys. Lett. \textbf{95}%
, 10008 (2011).

\bibitem{13} S. Banerjee, R. K. Gupta, A. Sen, J. High Energy Phys. 
\textbf{1103}, 147 (2011).

\bibitem{14} A. Sen, J. High Energy Phys. \textbf{1304}, 156 (2013).

\bibitem{15} A. Bagchi, R. Basu, J. High Energy Phys. \textbf{1403}, 020
(2014).

\bibitem{16} C. Keeler, F. Larsen, P. Lisbao, Phys. Rev. D \textbf{90}, 043011
(2014).

\bibitem{17} K. Nozari, A. Sefiedgar, Gen. Relativ. Gravit. \textbf{39},
501 (2007).



\bibitem{24} A. F. Ali, S. Das, E. C. Vagenas, Phys. Lett. B \textbf{678},
497 (2009).

\bibitem{25} J. Sadeghi, V. R. Shajiee, Eur. Phys. J. Plus \textbf{132},
132 (2017).

{\color{red}
\bibitem{Jizba} P. Jizba, H. Kleinert, F. Scardigli, Phys. Rev. D \textbf{81}, 084030 (2010).

\bibitem{Buoninfante} L. Buoninfante, G. G. Luciano, L. Petruzziello, Eur. Phys. J. C \textbf{79}, 663 (2019).

\bibitem{Chen} P. Chen, Y. C. Ong, D. -h. Yeom, J. High Energy Phys. \textbf{12}, 021 (2014).

\bibitem{Petruzziello} L. Petruzziello, F. Illuminati, Nature Commun. \textbf{12}, 4449 (2021).
}

\bibitem{26} S. Gangopadhyay, Int. J. Theor. Phys. \textbf{55}, 617 (2016).



\end{thebibliography}
\end{document}